\documentclass[twocolumn,showpacs,preprintnumbers,amsmath,amssymb,widetext]{revtex4}
\usepackage{graphicx}
\begin{document}
\title{Manipulating nonequilibrium magnetism through superconductors}

\author{Francesco Giazotto}
\email{giazotto@sns.it}
\author{Fabio Taddei}
\author{Rosario Fazio}
\author{Fabio Beltram}
\affiliation{NEST-INFM and Scuola Normale Superiore, I-56126 Pisa,
Italy}

\begin{abstract}
Electrostatic control of the magnetization of a normal mesoscopic conductor is analyzed in a hybrid superconductor-normal-superconductor system. This effect stems from the interplay between the non-equilibrium condition in the normal region and the Zeeman splitting of the quasiparticle density of states of the superconductor subjected to a static in-plane magnetic field. Unexpected spin-dependent effects such as magnetization suppression, \textit{diamagnetic-like} response of the susceptibility as well as spin-polarized current generation are the most remarkable features presented. The impact of scattering events is evaluated and let us show that this effect is compatible with realistic material properties and fabrication techniques.
\end{abstract}


\maketitle

The interplay between out-of-equilibrium transport and superconductivity \cite{kopnin} was recently 
successfully exploited in a number of systems in order to implement Josephson transistors~\cite{morpurgo,savin,giazotto}, 
$\pi$ junctions~\cite{Baselmans} and electron microrefrigerators~\cite{phystoday,pekola}, just to mention a few relevant examples. 
In this Letter we explore its potential in the area of magnetism \cite{yosida} and spintronics \cite{zutic} and present a novel approach to control the magnetization and spin-dependent properties of a mesoscopic normal conductor.
In particular, we show that manipulation of the (nonequilibrium) distribution of a normal metal through an applied voltage can lead to the control of a number of spin-dependent phenomena.
The key ingredients are superconductor electrodes 
(with energy gap $\Delta$) and a weak external magnetic field. 
The resulting Zeeman-split superconductor density of states (DOS) was originally exploited by Tedrow and Meservey to measure the spin-polarization of ferromagnets in the case of Al electrodes~\cite{tedrow1}.
As we shall argue, the interplay between Zeeman splitting and nonequilibrium yields dramatic consequences on quasiparticle dynamics stemming from the peculiar shape of the superconductor DOS whose energy gap compares well with magnetic fields readily accessible experimentally.

Let us consider the system sketched in Fig. 1. 
It consists of two superconducting reservoirs (S) connected by a 
mesoscopic normal metal wire (N) through tunnel contacts (I) of resistance $R_I$. The structure is biased at a voltage $V_C$ and in the presence of a static in-plane magnetic field $H$, applied either across the whole structure (Fig. 1(a), in the following referred to as \textbf{a}-type setup) or localized at the superconductors (Fig. 1(b), \textbf{b}-type setup). For the sake of simplicity let us assume a symmetric structure (a resistance asymmetry would not change the overall physical picture). As for the 
superconductors we focus on conventional low critical-temperature thin ($<$ 10 nm) films. In this case the effect of $H$ on the electron spin becomes dominant 
and, assuming negligible spin-orbit interaction \cite{tedrow0}, the superconductor DOS 
per spin is BCS-like but shifted by the Zeeman energy ($E_H=\mu_BH$), 
$\mathcal{N}^S_\sigma(\varepsilon)=\mathcal{N}^N_F|\text{Re}[(\varepsilon
+\sigma E_H)/2\sqrt{(\varepsilon+\sigma E_H)^2-\Delta^2}]|$ \cite{tedrow}, where 
$\varepsilon$ is the quasiparticle excitation energy measured from the Fermi energy ($\varepsilon_F$), $\mathcal{N}^N_F$ is the DOS in the normal state at  $\varepsilon_F$ (2 spin directions), $\mu_B$ is the Bohr magneton, and $\sigma=\pm 1$ refers to spin parallel(antiparallel) to the field. 

\begin{figure}[t!]
    \begin{center}
    \includegraphics[width=\columnwidth,clip]{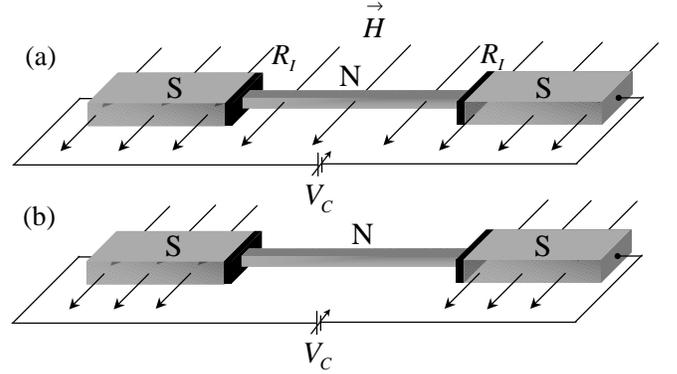}
    \end{center}
    \caption{Scheme of the structure investigated.
    An in-plane static magnetic field $H$ is applied across the whole SINIS system ((a), \textbf{a}-type setup) or localized at the S electrodes ((b), \textbf{b}-type setup).
    A finite voltage bias $V_C$ drives the
    normal metal out-of-equilibrium allowing
    to control its magnetization. The N wire is assumed quasi-one-dimensional.
    }
    \label{device}
\end{figure}

At a finite bias $V_C$, in the presence of $H$ and in the limit of negligible inelastic 
collisions, the steady-state distribution functions in the metal wire are 
spin-dependent and are given by~\cite{heslinga}
\begin{equation}
    f_{\sigma}(\varepsilon,V_C,H)
    =\frac{\mathcal{N}^L_\sigma\mathcal{F}^L+
    \mathcal{N}^R_\sigma\mathcal{F}^R}
    {\mathcal{N}^L_\sigma+
    \mathcal{N}^R_\sigma},
    \label{distr}
\end{equation}
where $\mathcal{F}^{L(R)}=f_0(\varepsilon\pm eV_C/2)$, $\mathcal{N}^L_\sigma=
\mathcal{N}^S_\sigma(\varepsilon+eV_C/2)$, $\mathcal{N}^
R_\sigma=\mathcal{N}^S_\sigma(\varepsilon-eV_C/2)$, $f_0(\varepsilon)$ is the Fermi distribution at lattice temperature $T$ and $e$ is the electron charge. 
Owing to the nonequilibrium regime driven by the applied electric field, the quasiparticle distributions corresponding to 
different spin species behave differently, $f_{+(-)}$ 
being shifted towards lower(higher) energy.
The magnetic properties of the N region are entirely determined by its (spin-dependent) 
quasiparticle distribution functions. 
The  magnetization density in the wire is indeed given by 
\begin{equation}
    \mathcal{M}(V_C,H)=\mu_B \int d\varepsilon\,
    [\mathcal{N}^N_+(\varepsilon)f_+(\varepsilon)-\mathcal{N}^N_-(\varepsilon)f_-(\varepsilon)],
\end{equation}
where $\mathcal{N}^N_\sigma(\varepsilon)=\frac{1}{2}\mathcal{N}^N(\varepsilon_F+\varepsilon+\sigma\mu_BH)$ and $\mathcal{N}^N(\varepsilon)$ is the N region DOS in the absence of magnetic field. 
The function $\mathcal{M}(V_C,H)$ vs $V_C$ is displayed in Fig. 2(a,b) for different magnetic-field values. We assumed  a silver (Ag)  N region (with $\mathcal{N}^N_F=1.03\times10^{47}$ J$^{-1}$m$^{-3}$)  at temperature $T=0.1\,T_c$, 
where $T_c=(1.76\,k_B)^{-1}\Delta=1.196$ K is the critical temperature of bulk aluminum (Al, the material forming the S regions) and $k_B$ is the Boltzmann constant.  
When $H$ is applied across the whole SINIS structure (\textbf{a}-type setup), $\mathcal{M}$ decreases upon increasing $V_C$ starting from its equilibrium value $\mathcal{M}_{Pauli}=\mu_B^2\mathcal{N}_F^NH$ typical of a Pauli paramagnet \cite{yosida} (see Fig. 2(a)).
$\mathcal{M}$ shows a complete suppression for $V_C \gtrsim \Delta/e$, i.e. the N region is \emph{demagnetized}.
The inset of Fig. 2(a) shows how $\mathcal{M}(V_C)$ is weakly dependent on the lattice temperature up to $T=0.4T_c$ owing to the BCS $\Delta(T)$ dependence together with the temperature-induced broadening of $f_0(\varepsilon)$.
Conversely, when the magnetic field is localized at the S electrodes (\textbf{b}-type setup) a  \emph{negative} magnetization is induced in the wire (see Fig. 2(b)).
Note that $\mathcal{M}$ is antiparallel to $H$. Therefore, the N region behaves as a "diamagnet" \cite{newnota}. 
For $eV_C\gtrsim\Delta$ the wire susceptibility $\chi$ (shown in Fig. 1(c) at $T=0.1\,T_c$) reaches 
the Pauli value but with opposite sign $\chi=\partial \mathcal{M}/\partial H=-\mu_B^2 \mathcal{N}^N_F=-\chi_{Pauli}$. This gives rise to a sort of "artificial" Pauli diamagnetism. 

\begin{figure}[t!]
    \begin{center}
    \includegraphics[width=\columnwidth,clip]{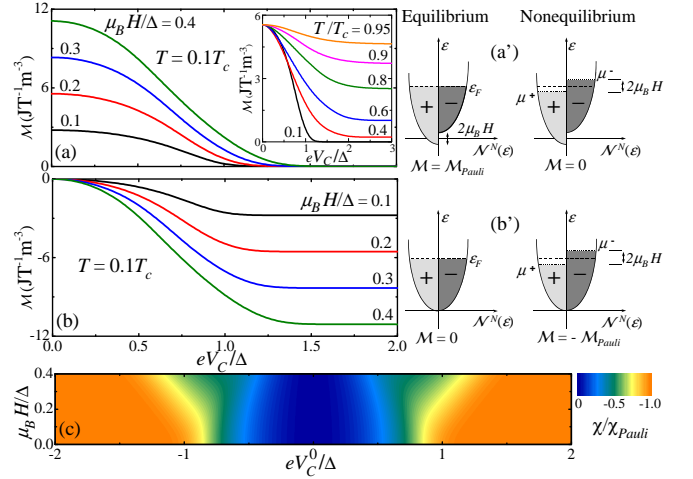}
    \end{center}
    \caption{(color) (a) Magnetization density $\mathcal{M}$ vs 
    bias voltage  $V_C$ at $T=0.1\,T_c$ for different magnetic fields ($H$) for \textbf{a}-type setup (see Fig. 1(a)). Inset: $\mathcal{M}$ vs $V_C$ for different temperatures at $H=0.2\,\Delta/\mu_B$.
    (b) The same as in (a) for \textbf{b}-type setup. (a') Schematic diagrams of the N region density of states and quasiparticle occupation both at equilibrium (left) and nonequilibrium (right) for \textbf{a}-type setup. (b') The same as in (a') for \textbf{b}-type setup. (c) Contour 			plot of the normalized magnetic susceptibility $\chi/\chi_{Pauli}$ vs 
    $V_C$ and $H$ at $T=0.1\,T_c$ for \textbf{b}-type setup.}
\end{figure}

Insight into the physical origin of this superconductivity-controlled magnetism can be qualitatively gained by considering the (zero-temperature) steady-state DOS diagrams of Fig. 2(a',b'), where the normal metal is described by parabolic subbands typical of a free electronlike paramagnetic conductor such as silver.
At \emph{equilibrium} the occupation of quasiparticle states is identical for both spin species leading to $\mathcal{M}=\mathcal{M}_{Pauli}$ and $\mathcal{M}=0$ for \textbf{a}-type and \textbf{b}-type setups, respectively.
When $V_C\neq0$, electron distributions for the two spin populations are characterized by distinct chemical potentials  $\mu^\sigma$. Since $\mu^+ < \mu^-$ the occupation 
of spin states antiparallel to the magnetic field is favored with respect to the parallel one, owing to the opposite energy shift of the superconductor spin-dependent 
DOS in the external magnetic field. This leads to a reduction of $\mathcal{M}$ for the \textbf{a}-type setup and to negative magnetization for the \textbf{b}-type setup. 
In particular, at $V_C\sim \Delta/e$ the chemical potential separation is $\delta\mu=\mu^+-\mu^-\sim-2\mu_BH$ for both setups.
This shows a full \emph{electrostatic} control of the magnetization, a unique feature of the present system \cite{nota1}. 
The superconducting reservoirs are \emph{essential} elements for these effects to be present and replacing them by ordinary normal contacts would only always lead to a negligible paramagnetic correction to the actual $\mathcal{M}$ (second order in the small quantities $\mu_BH/\varepsilon_F$ and $eV_C/\varepsilon_F$).

The experimental accessibility of this operational principle must be carefully assessed. 
Electrons in metals experience both elastic and 
inelastic collisions. 
The latter drive the system to equilibrium and can be expected to hinder the observation of the phenomena discussed here. Our analysis will show a remarkable robustness of these effects. 
At low temperatures (typically below 1 K) electron-electron scattering \cite{alt}, and scattering with magnetic impurities \cite{kaminski,anthore} are the dominant sources of 
inelastic collisions \cite{pothier,anthore,nagaev}. 
Since $R_I$ is in general large compared to wire resistance ($R_N= L/\mathcal{N}_F^Ne^2DA$, where $L$ is the length, $A$  the cross-section and $D$ the wire diffusion constant), we can assume that $f_{\sigma}$ changes only at the interfaces, being essentially constant elsewhere \cite{SINIS}.
The effect of electron-electron scattering due to direct Coulomb interaction on the spin-dependent distributions can be accounted for by solving a pair of coupled stationary kinetic equations: 
\begin{equation}\label{kinetic}
\left\{ 
\begin{split}
D\frac{\partial^2f_+(\varepsilon)}{\partial x^2}={\cal I}_{\text{coll}}^+(\varepsilon)\\
D\frac{\partial^2f_-(\varepsilon)}{\partial x^2}={\cal I}_{\text{coll}}^-(\varepsilon),
\end{split}
\right. 
\end{equation}
together with the Kuprianov-Lukichev boundary conditions at the NIS interfaces \cite{KL}.
In (\ref{kinetic}) ${\cal I}_{\text{coll}}^{\sigma}(\varepsilon)$ is the net collision rate at energy $\varepsilon$, functional of the distributions functions $f_\sigma$, defined by 
\begin{equation}
{\cal I}_{\text{coll}}^{\sigma}(\varepsilon)={\cal I}_{\text{coll}}^{\text{in}\sigma}(\varepsilon)-{\cal I}_{\text{coll}}^{\text{out}\sigma}(\varepsilon),
\end{equation}
where
\begin{equation}\label{Icollin}
\begin{split}
{\cal I}_{\text{coll}}^{\text{in}\sigma}(\varepsilon)=
[1-f_{\sigma}(\varepsilon)]\int d\omega \frac{k(\omega)}{2}f_{\sigma}(\varepsilon-\omega)\\
\int dE \left\{ f_+(E+\omega)[1-f_+(E)]+
f_-(E+\omega)[1-f_-(E)]
\right\}
\end{split}
\end{equation}
and
\begin{equation}\label{Icollout}
\begin{split}
{\cal I}_{\text{coll}}^{\text{out}\sigma}(\varepsilon)=
f_{\sigma}(\varepsilon)\int d\omega \frac{k(\omega)}{2}[1-f_{\sigma}(\varepsilon-\omega)]\\
\int dE \left\{ f_+(E)[1-f_+(E+\omega)]+
f_-(E)[1-f_-(E+\omega)]
\right\} .
\end{split}
\end{equation}
In (\ref{Icollin}) and (\ref{Icollout}), $k(\omega)=\kappa_{ee} \omega^{-3/2}$ according to the theory of screened Coulomb interaction \cite{alt2} for a quasi-one dimensional wire, where $\kappa_{ee}=(\pi\sqrt{2D}\hbar^{3/2}\mathcal{N}_F^NA)^{-1}$ \cite{kamenev,huard}. By making (\ref{kinetic}) dimensionless \cite{SINIS}, the strength of the electron-electron interaction can then be expressed as $\mathcal{K}_{coll}=(R_I/R_N)(L^2 \kappa_{ee} /D)\sqrt{\Delta}=(L/\sqrt{2})(R_I/R_K)\sqrt{\Delta/\hbar D}$, where $R_K=h/2e^2$. 

We analyzed quantitatively a realistic Ag/Al SINIS microstructure \cite{phystoday,pekola} with $L=1\,\mu$m, $A=0.2\times0.02\,\mu$m$^2$, and $R_I=10^3 \,\Omega$. 
\begin{figure}[t!]
    \begin{center}
    \includegraphics[width=\columnwidth,clip]{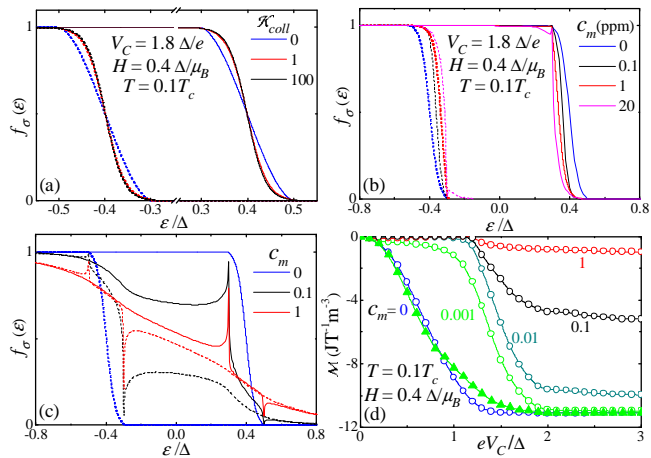}
    \end{center}
    \caption{(color)
    (a) 
    Spin-dependent distribution functions $f_\sigma(\varepsilon)$  vs energy $\varepsilon$ calculated for three $\mathcal{K}_{coll}$ values at $H=0.4\,\Delta/\mu_B$, $V_C=1.8\,\Delta/e$ and $T=0.1\,T_c$. Solid(dashed) lines correspond to antiparallel(parallel) spin species.
    (b) 
    $f_\sigma(\varepsilon)$ vs $\varepsilon$ calculated for various $c_{m}$ values at $H=0.4\,\Delta/\mu_B$, $V_C=1.8\,\Delta/e$ and $T=0.1\,T_c$ for \textbf{a}-type setup. 
    (c)
    The same as in (b) for \textbf{b}-type setup. 
    (d)
    Magnetization density $\mathcal{M}$ vs  $V_C$ at $H=0.4\,\Delta/\mu_B$ and $T=0.1\,T_c$ for different magnetic impurity concentration. Open circles refer to \textbf{b}-type setup, filled triangles to \textbf{a}-type setup for $c_m =0.001$. The latter were shifted  by $-\mathcal{M}_{Pauli}$. Data in (b)-(d) were obtained assuming $D=0.02$ m$^2$s$^{-1}$, $T_K=40$ mK and $S=\frac{1}{2}$.}
\end{figure}
Figure 3(a) illustrates the effect of electron-electron scattering. We solved (\ref{kinetic}) with $H=0.4\,\Delta/\mu_B$, $V_C=1.8\,\Delta/e$ and $T=0.1\,T_c$ for several $\mathcal{K}_{coll}$ values from negligible ($\mathcal{K}_{coll}=0$, blue lines), to
moderate ($\mathcal{K}_{coll}=1$, red lines) and extreme ($\mathcal{K}_{coll}=100$, black lines) \cite{alt2,kamenev}. 
As expected, electron-electron interactions have virtually no impact.
By increasing the strength of Coulomb interaction 
the quasiparticle distribution of each spin species relaxes toward 
spin-dependent 
Fermi functions still characterized by different chemical potentials (a similar effect is expected in the presence of interaction with the lattice phonons \cite{Roukes}). 
As a result the nonequilibrium magnetization in the normal wire here presented is virtually unaffected. 

The situation drastically changes if we assume the presence of magnetic impurities in the N region, due to the resulting spin-flip processes. 
Above the Kondo temperature ($T_K$), the distribution functions can be calculated including,  in the right-hand-side of Eqs. (\ref{kinetic}), an additional term derived by generalizing the theory developed by G\"{o}ppert and Grabert in Ref. \cite{goppert}. 
It is noteworthy to mention that its strength is proportional, apart from the electron and magnetic impurity spin coupling constant, to the \textit{total} number of magnetic impurities present within the wire (i.e., to the product $c_{m}LA$, with $c_{m}$ the impurity concentration) and $R_I$.
The resulting distribution functions relative to the \textbf{a}-type setup are shown in Fig. 3(b) at $H=0.4\,\Delta/\mu_B$, $V_C=1.8\,\Delta/e$ and $T=0.1\,T_c\approx 120$ mK for various $c_{m}$ values expressed in parts per million (ppm).
We assumed $D=0.02$ m$^2$s$^{-1}$ (typical of high-purity Ag), magnetic impurities with spin $S=\frac{1}{2}$, and $T_K=40$ mK (as appropriate, for example, for Mn impurities in Ag) \cite{coles,pierre}. 
By increasing $c_{m}$, examination of the figure immediately shows that spin-dependent distributions are marginally affected even for impurity concentrations as large as 20 ppm. This shows that in the \textbf{a}-type setup the nonequilibrium $\mathcal{M}$ is relatively \emph{insensitive} to large amounts of magnetic impurities. 
Figure 3(c) shows the $f_{\sigma}(\varepsilon)$ calculated for various $c_{m}$ values for the \textbf{b}-type setup.
In such a case, by contrast, the spin-dependent distribution functions tend to merge for much lower values of $c_{m}$ thus suppressing the induced magnetization. In the presence of a magnetic field across the N region (\textbf{a}-type setup) impurity spins tend to polarize yielding a suppression of spin-flip relaxation processes for the field intensities  of interest here \cite{kaminski,goppert,goppert2,anthore}. This does not occur in the \textbf{b}-type setup and makes magnetic impurities more effective in mixing spins.
The full behavior of $\mathcal{M}(V_C)$ for \textbf{b}-type setup at $T=0.1\,T_c$ and $H=0.4\,\Delta/\mu_B$ is displayed in Fig. 3(d) for several $c_{m}$ values (open circles). For comparison, $\mathcal{M}(V_C)$ for \textbf{a}-type setup (filled triangles) is shown at low impurity concentration. We wish to underline the robustness of the induced magnetization, $\mathcal{M}$ being suppressed only for rather large concentrations: the latter can in fact be limited to less than 0.01 ppm in currently available high-purity metals \cite{pierre}.

\begin{figure}[t!]
    \begin{center}
    \includegraphics[width=\columnwidth,clip]{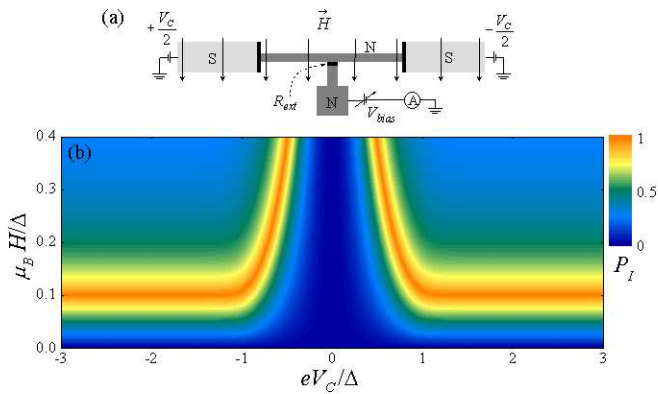}
    \end{center}
    \caption{(color)
    (a) Scheme of the spin-polarized current source in the presence of a uniform $H$. Spin-polarized current can be extracted by biasing the middle terminal with
    $V_{bias}$. 
    (b) Contour plot of the nonequilibrium current polarization $P_I$ vs  $V_C$ 
    and $H$ for $V_{bias}=0.1\,\Delta/e$ at $T=0.1\,T_c$. 
    }
    \label{currentpol}
\end{figure}

These results on the robustness of these effects in realistic structures make it appropriate to investigate their potential for device implementation. An immediate area of application is \emph{spintronics} \cite{zutic}. Let us consider a microstructure like that shown in Fig. 4(a), where a normal "probe" terminal is tunnel-coupled to the wire \cite{nota3}. 
Upon voltage biasing with $V_{bias}$, the presence of spin-dependent distributions
in the N region yields a finite current polarization $P_I$ defined as
\begin{equation}
    P_I(V_C,H,V_{bias})=
    \frac{|I_-|-|I_+|}{|I_-|+|I_+|}~,
\end{equation}
where 
\begin{equation}
\begin{split}
I_\sigma(V_C,H,V_{bias})=\wp\int d\varepsilon\,\mathcal{N}_{\sigma}^N(\varepsilon)\mathcal{N}_{\sigma}^N(\varepsilon+eV_{bias})\\
\times[f_\sigma(\varepsilon,V_C,H)-f_0(\varepsilon+eV_{bias})]
\end{split}
\end{equation}
is the spin-dependent current flowing through the additional terminal. Furthermore, $\wp=[2e(\mathcal{N}_F^N)^2R_{ext}]^{-1}$ and $R_{ext}\gg R_I$ is the probe junction resistance. The calculated nonequilibrium $P_I$ is displayed in Fig. 4(b) for $T=0.1\,T_c$ and  $V_{bias}=0.1\,\Delta/e$ as a function of $V_C$ and $H$. 
We emphasize that for easily attainable values of $V_C$ and $H$ a 100\% 
spin-polarized current consisting of the \textit{antiparallel} spin species can be achieved. A quantitative estimate for realistic parameters shows that sizable spin-polarized currents can be available. For example, assuming $R_{ext}=10^4$ $\Omega$ at $T=0.1\,T_c \approx 120$ mK and for an external field of 1.5 T the fully spin-polarized current reaches values up to about $10^{-8}$ A.
We stress that $P_I$ values largely exceed 50\% 
over a wide region in the ($V_C,H$) plane. Note that at equilibrium by placing the wire in external magnetic fields of comparable intensity only values of $P_I$ of the order of $10^{-6}\div10^{-5}$ would be obtained. 

In conclusion, we have presented a scheme to control the magnetic properties of a mesoscopic metal. Magnetism suppression as well as artificial Pauli diamagnetism can be accessed in metal-superconductor microstructures thus making available a number of characteristics of much relevance in light of possible applications: (1) Generation of $100\%$ spin-polarized currents without invoking the use of magnetic materials; (2) \emph{full-electrostatic} 
control of magnetization over complex nanostructured metallic arrays for enhanced
performance and optimized device geometries; (3) reduced power dissipation ($10^{-14}
\div 10^{-11}$ W depending on the control voltage) owing to the very small driving currents 
intrinsic to SIN junctions; (4) high magnetization 
switching frequencies up to $10^{11}$ Hz \cite{SINIS}; (5) ease of fabrication that can take advantage of the well-established metal-based tunnel junction technology.

This work was supported in part by MIUR under FIRB "Nanotechnologies and nanodevices for information society" contract RBNE01FSWY and by RTN-Spintronics.


\begin{references}
\bibitem{newnota} It is worth mentioning that the induced magnetization vanishes if the external fields acting on the two superconductors are arranged antiparallel with each other.
\bibitem{nota1} Note that one superconducting electrode is sufficient in order to control the wire magnetization.
\bibitem{nota3}
        A similar scheme would also allow an accurate determination of $f_{\sigma}(\varepsilon)$ from the voltage-dependent differential conductance of the probe junction \cite{anthore}.
\end{references}
\end{document}